\documentclass{aa}
\usepackage{graphicx}
\def\b0048{\object{PKS\,B0048$-$097}}

\begin{document}

\title{A Quasi-Periodic Modulation of the Radio Light Curve of the Blazar \b0048}
\titlerunning{Quasi-Periodic Modulation of PKS\,B0048$-$097}
   \author{M. Kadler\inst{1}$^{\rm ,}$\inst{2} 
          \and
          P. A. Hughes\inst{3}
          \and
          E. Ros\inst{1}
          \and
          M. F. Aller\inst{3}
          \and
          H. D. Aller\inst{3}
          }

   \offprints{mkadler@milkyway.gsfc.nasa.gov}

   \institute{Max-Planck-Institut f\"ur Radioastronomie,
              Auf dem H\"ugel 69, 53121 Bonn, Germany\\
              \email{mkadler@milkyway.gsfc.nasa.gov, ros@mpifr-bonn.mpg.de}
         \and
              Exploration of the Universe Division, NASA Goddard Space Flight Center, Greenbelt Road, Greenbelt, MD 20771, USA
         \and
              Astronomy Department, University of Michigan,
                Ann Arbor, MI 48109-1042, USA\\
                \email{[phughes,mfa,haller]@umich.edu}
             }

   \date{Received 21 November 2005; Accepted 23 May 2006}

   \abstract{ 
In this letter we present the results of a wavelet
analysis of the radio light curve of the BL\,Lac Object PKS\,B0048$-$097 from 
the University of Michigan Radio Astronomy Observatory monitoring
program at 8\,GHz
during twenty-five years, from 1979 to 2004.
The results show a remarkable periodicity of 450--470 days in the early 1980s 
that changed to a $\sim$585 day periodicity in the late 1980s to early 1990s.
A less pronounced $\sim$400 day periodicity is found after $\sim 1995$.
Very-long-baseline interferometry imaging at 15\,GHz shows
dramatic structural changes in the usually unresolved source
between two epochs, 1995.57 and 2002.38. The pronounced northward
directed jet seen in the 2002 image differs by more than 90$^\circ$ 
in direction from the source structure found in earlier epochs.
These findings make PKS\,B0048$-$097 a
primary target for multi-wavelength observations and intensive
radio monitoring to decipher the blazar-variability phenomenon.
   }

   \maketitle
%

\section{Introduction}

Combined multi-frequency (radio to X-ray/$\gamma$-ray) observations
of active galactic nuclei (AGN) can yield important insights into the physics
of relativistic outflows associated with super-massive black holes. 
In particular the emission of blazars (BL\,Lac objects and OVV quasars, see
Ulrich, Maraschi \& Urry \cite{Ulr97} for a description on the nature of
blazars) is believed
to be dominated over the whole electromagnetic spectrum by the most compact 
regions of relativistic jets.
{Blazars exhibiting 
quasi-periodic behaviour}
in principle allow the inter-relation between the compact radio jet
and the broadband spectral energy distribution to be deciphered
via quasi-simultaneous multi-frequency observations,
especially if the time scales are small enough to consider variability data over
multiple cycles,
but such objects
are rare. 

Hughes, Aller, \& Aller (\cite{Hug98}) find a persistent 
modulation of the total flux and polarisation of the BL\,Lac object \object{OJ\,287}.
The dominating period of $\sim$1.12\,yr in the 1980s was later displaced by a
strong
$\sim$1.66\,yr {periodicity during the 1990s}. The relationship between these two variations 
was interpreted by Hughes, Aller, \& Aller (\cite{Hug98}) in terms of a 
``shock-in-jet'' model while the complex periodic optical variability 
of OJ\,287 (period between major outbursts $\sim 11.6$\,yr),
has been interpreted, e.g., in terms of a ``precessing-jet'' model by 
Abraham (\cite{Abr00}).
Stirling et al. (\cite{Sti03}) suggest an oscillating ``nozzle'' structure
of the inner jet of \object{BL\,Lac}
with a period of $\sim 2$ years, based on radio millimetre
flux-density monitoring and VLBI observations between 1998 and 2001
(see also Mutel \& Denn \cite{Mut05}).
Villata et al. (\cite{Vil04}) analyse the optical and radio long-term
variability of BL\,Lac and find a gradually lengthening $\sim 8$\,yr periodicity.  
The optical and radio long-term variability of the BL\,Lac object \object{AO\,0235+16}
has been investigated by Raiteri et al. (\cite{Rai01}) revealing a possible
5--6 year periodicity based on long-term monitoring data, in the radio
regime particularly on data of  
the 
UMRAO\footnote{University of Michigan Radio Astronomy Observatory; see
{\tt http://www.astro.lsa.umich.edu/obs/radiotel/umrao.html}} 
{database}. Ostorero, Villata \& Raiteri  (\cite{Ost04}) 
applied the helical jet model of Villata \& Raiteri (\cite{Vil99})
to the case of AO 0235+16, 
interpreting the data in terms of Doppler factor variations due to 
changes of the viewing angle because of the helix rotation.
In all these cases, attempts have been made to understand the periodic
behaviour as a result of orbital motion 
of two black holes in a binary system,
helical jet structures, shocks, and instabilities of the disk or jet-plasma flow.

In this letter, we report on the detection of a particularly strong,
and persistent quasi-periodic {long-term} modulation 
of the compact-jet emission of the BL\,Lac object 
PKS\,B0048$-$097. The 
comparatively short time scale of $\sim$350\,days to $\sim$600\,days 
and the large amplitude of
the variability make PKS\,B0048$-$097 a particularly well-suited source for future multi-frequency
campaigns on the broadband spectral emission of blazar sources and tests of
blazar-periodicity models. 

In Sect.~2, we present the observational data and their analysis.  
We discuss the results of a cross-wavelet analysis of 
twenty five years of the University of Michigan Radio Astronomy Observatory
(UMRAO) monitoring
of PKS\,B0048$-$097 and 
discuss {its parsec-scale structure} during two epochs observed as part of the
Very Long Baseline Array (VLBA) 2\,cm Survey. 
Finally, we summarize our results and their implications
for future blazar studies, 
in Sect.~3.  

\section{Background, Observations and Analysis}

PKS\,B0048$-$097 is a BL\,Lac object with unknown redshift.  A lower limit of $z=0.5$ 
is postulated by Falomo (\cite{Fal96}) from \textsl{HST} data.  
The flux density of PKS\,B0048$-$097 has been measured since 1979 
at 4.8\,GHz, 8.0\,GHz, and
14.5\,GHz using the University of Michigan 
26\,m paraboloid.
The source is observed once every three months at all three frequencies
as part of the UMRAO BL\,Lac Observing Program
(Aller et al. \cite{All99}; Aller, Aller \& Hughes \cite{All03}).
Its light curve is best sampled at 8.0\,GHz 
(see Fig. \ref{fig:morlet}) exhibiting pronounced maxima
during several well defined time periods.


\subsection{Wavelet Analysis}
We used a continuous wavelet analysis, which quantifies the behaviour of a signal on
different temporal scales,
to investigate the available UMRAO data for periodicities. As a function of time, the signal is convolved
with a localised wave-packet, that is translated along the series,
for a number of `dilations' of the wave-packet 
({Hughes et al.\ } \cite{Hug98}).  
This technique has the great advantage
of preserving temporal locality: a gap in the time series will be evident
along the corresponding line in transform space, and events that are distinct
in the signal will have distinct counterparts in transform space.  A Morlet
wavelet is particularly suited to the analysis of time series, as it is
complex, so the real part of the transform exhibits an oscillatory behaviour
corresponding to periodicity in the time series, while the modulus provides a
measure of the power in different components of the signal.

\begin{figure}[bht]
\centering
\includegraphics[width=\columnwidth]{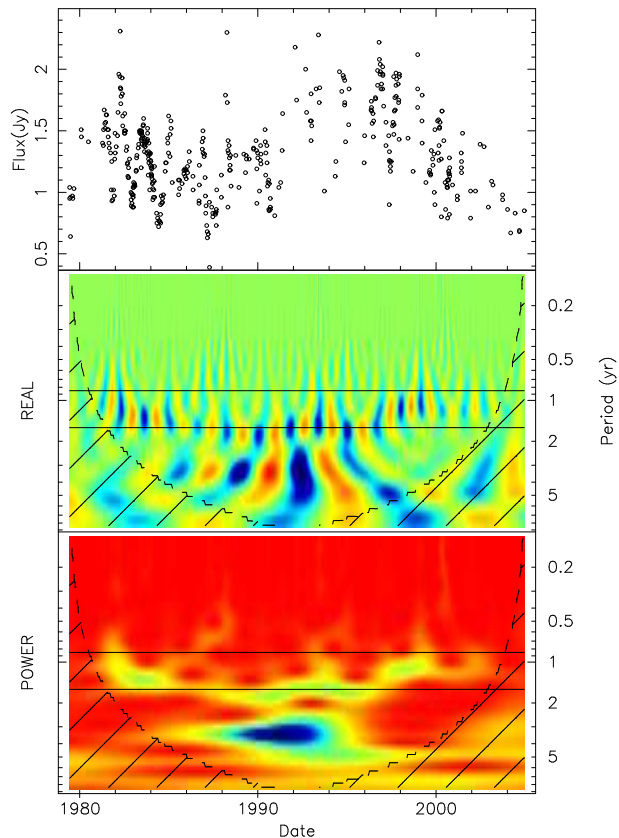}
\caption{A continuous Morlet { wavelet} transform of the time series 
data 
{of the flux density for PKS\,B0048$-$097 taken at the UMRAO at 8.0\,GHz}. 
The panels show the {flux density} time series (top), the real part of the transform (middle)
and the modulus (power) of the transform (bottom). 
The real part displays clearly the signature
of a quasi-periodic component with a slowly varying $\sim$ 1--2\,yr period.
Power at this dilation extends
across the entire series. The horizontal lines in the middle and
bottom panel enclose
the 95\,\% significance region of the detected periodicity determined by
a cross-wavelet transform analysis (see text).
The hatched areas of the transform panels show the cone of influence,
    where edge effects resulting from the finite length of the time series
    decrease the wavelet coefficient power by more than a factor $e^2$.
\label{fig:morlet}
}
\end{figure}

Such an analysis has been applied to the time series for 
PKS\,B0048$-$097, and,
as seen in Fig.~\ref{fig:morlet}, a distinct pattern is seen in both the real part of the
source wavelet transform, and its modulus, corresponding to a quasi-periodic
component containing a modest fraction of the overall power, and masked in a
Fourier power spectrum, because of drifts in time scale across the data
window.  
In the cross wavelet transform technique, described, e.g., by 
Kelly et al. (\cite{Kel03}), the continuous transform of the signal is 
convolved with a set of template periodic signals.
This analysis 
has been used to
quantify the result (apparent on visual inspection of 
the top panel at Fig.~\ref{fig:morlet}), 
establishing a
time scale of 385--470 days
with a confidence of $>99.9$\% that this component
does not arise by chance from random patterns associated with a lag-1
autoregressive process. 

\begin{figure}[h!]
\centering
\includegraphics[clip, width=\columnwidth]{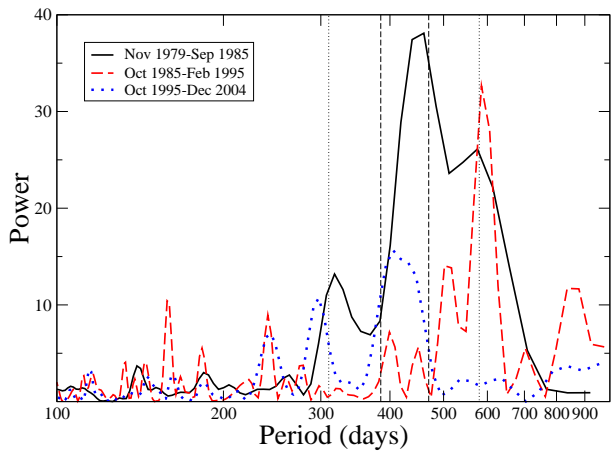}
\caption[Lomb-Scargle periodogram for the flux density light curve of PKS\,B0048$-$097]{Lomb-Scargle periodogram for three different parts of the
time series of UMRAO PKS\,B0048$-$097 8\,GHz monitoring data.
The dotted and dashed vertical lines show the 95\,\% and 99.9\,\% significance lines,
respectively, derived from the cross wavelet transform of the full
light curve.
\label{fig:scargle}}
\end{figure}                    

The varying time scale of the disclosed periodicity has been further investigated
using a Lomb-Scargle periodogram analysis
(Lomb \cite{Lom76}, Scargle \cite{Sca82}).
In a L-S periodogram, the classical discrete Fourier periodogram
is redefined in such a manner to as to make it invariant to a shift of the
origin of time. 
The L-S periodogram analysis has been performed 
using the {\sc period}
software package\footnote{The {\sc period} software package is provided
by the Starlink Project which was run by CCLRC on behalf of PPARC.
See {\tt \small http://star-www.rl.ac.uk}}.
The L-S analysis
between November 1979 and September 1985 shows maximal power at
$\sim 400$--$460$ days and a secondary signal at $\sim 580$ days
(see Fig.~\ref{fig:scargle}).
Between October 1985 and February 1995, only one narrow peak
at $\sim 585$ days is present. After October 1995, the highest power
is found at a time scale of $\sim 405$ days, that is considerably less
pronounced than the periodicities at the earlier time ranges.
The significance of the individual peaks in Fig.~\ref{fig:scargle} is difficult to
quantify because of the not {\it a-priori} known frequency dependence of the
underlying noise process\footnote{Only the assumption of a white noise process would allow a
  simple significance threshold parallel to the frequency axis to be placed. 
Admitting
  a red noise process would require us to consider a range of possible
  power laws for the process, which would lead to a range of significance
  areas (not parallel to the frequency axis).
Note, that the formal L-S false-alarm-probability lies below 1\,\%
(with 95\,\% confidence) for all major peaks in Fig.~\ref{fig:scargle}}. 
We base our detection on the highly-significant 
signal in the global power spectrum.
Figure~\ref{fig:scargle} illustrates that over the full $\sim 25$ years of
UMRAO 8\,GHz radio light curve monitoring of PKS\,B0048$-$097
the maximal power found by 
the L-S analysis
is located 
within the significance ranges found by the 
cross-wavelet analysis, but that its
time scale and power vary with time.

\subsection{Structural Variability}
\begin{figure}
\centering
\includegraphics[width=\columnwidth]{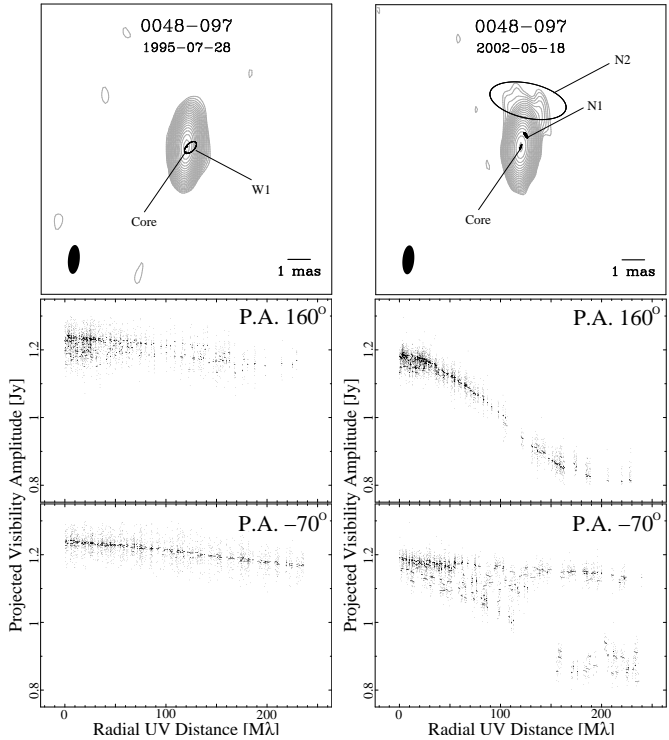}
\caption{{VLBI results of PKS\,B0048$-$097 at 15\,GHz.}
Top Panels: Naturally weighted images of
PKS\,B0048$-$097 in 1995.58 (left) and in 2002.38 (right)
at $\lambda 2$\,cm obtained from VLBA 2\,cm Survey
observations.
{Image parameters are given in Table~\ref{vlbi_maps}.}
A jet towards the north is visible in the 2002 image while a slight extention to the west
is seen in the earlier 1995 image.
Best fitting Gaussian-model-component approximations to the $(u,v)$-data are indicated
as ellipses.
{Middle and bottom panels:}
The projected visibility along P.A. $160^\circ$ and $-70^\circ$ for both
epochs.
Clear deviations from a flat
point-source visibility indicate an elongation of the
source along different position angles.}
\label{fig:0048--097}
\end{figure}

On parsec scales, Shen et al. (\cite{She97}) and 
Gabuzda, Pushkarev \& Cawthorne  (\cite{Gab99}) 
both report a core-jet morphology
of PKS\,B0048$-$097, however the reported jets differ by $\sim$40 degrees
in position angle (P.A.). While
Shen et al. (\cite{She97}) report a P.A. of $\sim -160^\circ$ (south-westward) in 
epoch 1992.9
{from 5\,GHz VLBI observations,} 
Gabuzda et al. (\cite{Gab99})
find a jet at P.A. $\sim +160^\circ$ (south-eastward) in epoch 1992.2
{also from 5\,GHz data.}  In both cases the $(u,v)$-coverage
was not optimum.
In the VLBA 2\,cm Survey observations before 2002, PKS\,B0048$-$097 showed no
clear resolved structure.
Figure~\ref{fig:0048--097} shows the milliarcsecond structure of PKS\,B0048$-$097
in epochs 1995.57 and 2002.38 (compare Table~\ref{vlbi_maps}).
In 1995.57, a weak westward directed jet is found, partially resolved only by the longest
east-west baselines.
The $(u,v)$-data obtained in 2002.38
show clear evidence for a resolved north-south structure along
P.A. $\sim -30^\circ$ with the jet pointing in a direction more
than 90$^\circ$ different from what was reported
by Shen et al. \cite{She97} and Gabuzda et al. (\cite{Gab99}). 
{Considering these discrepancies between different works,
PKS\,B0048$-$097} might represent a case of extreme jet-ejection-angle variation.
This would be of particular interest in conjunction with  a putative
periodicity of the radio light curve as revealed from the UMRAO data above.
In the scenario of a precessing jet, PKS\,B0048$-$097 might represent a highly attractive target
to study the broadband jet emission of a BL\,Lac object  at different angles
to the line of sight.

On the basis of the available data, it is not possible to derive the time
scales of the structural variability. VLBA 2\,cm Survey observations of
PKS\,B0048$-$097 have been conducted during two
additional epochs between 1995 and 2002 but the data quality in both cases does 
not allow us to compare the milliarcsecond structure directly to the two
images presented here, due to shorter integration times and sub-optimal
$(u,v)$-coverage. After 2002, the source has been observed several times
as part of the VLBA 2\,cm Survey continuation project, MOJAVE. An analysis of
these observations, which provide full polarimetric information, is currently
being performed. 
{A first MOJAVE image of PKS\,B0048$-$097 in 2003 showing a northward-directed jet similar
to the 2002 source structure can be found in Lister \& Homan (\cite{Lis05}).}

   \begin{table}
      \caption[]{VLBI Results}
         \label{vlbi_maps}
\resizebox{\columnwidth}{!}{%
         \begin{tabular}{@{}cc@{~}c@{~}cc@{~}c@{~}c@{~}c@{}}
            \hline
            \noalign{\smallskip}
& \multicolumn{3}{c}{Image parameters$^{\mathrm{a}}$} & \multicolumn{4}{c}{Model fit parameters$^{\mathrm{b}}$}\\
            Epoch      &  Beam$^{\mathrm{c}}$\ & $S_{\rm peak}$$^{\mathrm{d}}$ & rms$^{\mathrm{e}}$ & Comp.$^{\mathrm{f}}$\ & $r$$^{\mathrm{g}}$ & P.A.$^{\mathrm{h}}$ & $S$$^{\mathrm{i}}$ \\
                       & {\footnotesize [(mas$\times$mas),$^\circ$]} & {\footnotesize [Jy/beam]} & {\footnotesize [mJy/beam]} &  & {\footnotesize [mas]} & {\footnotesize [$^\circ$]} & {\footnotesize [Jy/beam]} \\
            \noalign{\smallskip}
            \hline
            \noalign{\smallskip}
1995-07-28 & $(1.3 \times 0.5)$; $-5.2$ & $1.21$ & $0.3$ & Core & $0$ & --- & $1.186$ \\
           &                            &        &       & W1   & $0.2$ & $-97.4$ & $0.052$ \\
            \noalign{\smallskip}
            \hline
            \noalign{\smallskip}
2002-05-18 & $(1.3 \times 0.5)$; $-2.6$ & $1.09$ & $0.2$ & Core & $0$ & --- & $1.074$ \\
           &                            &        &       & N1   & $0.6$ & $-22.2$ & $0.110$ \\
           &                            &        &       & N2   & $2.1$ & $8.7$ & $0.012$ \\
            \noalign{\smallskip}
            \hline
\noalign{\smallskip}
         \end{tabular}
}
\begin{scriptsize}
$^{\mathrm{a}}$ The lowest contour is 1\,mJy/beam in both images, contours increase by factors of $\sqrt{2}$; 
$^{\mathrm{b}}$ Formal uncertainties in the model fits are too small 
and therefore are not printed in the table;
$^{\mathrm{c}}$ Size and orientation of the restoring beam;
$^{\mathrm{d}}$ Peak flux density per beam area;
$^{\mathrm{e}}$ Noise level;
$^{\mathrm{f}}$ Individual Gaussian model-fit component;
$^{\mathrm{g}}$ Distance from core component;
$^{\mathrm{h}}$ Position angle of model-fit component;
$^{\mathrm{i}}$ Flux density of model-fit component per beam area.
\end{scriptsize}
   \end{table}

\section{Summary and Implications}

The BL\,Lac object PKS\,B0048$-$097 exhibits strong structural variability 
on sub-milliarcsecond scales and pronounced
radio, optical (see, e.g., Pica et al. \cite{Pic88}) and X-ray flux variability.
We have demonstrated that the radio light curve of PKS\,B0048$-$097 
has a strong modulation with a period of $\sim 350$\,days
to  $\sim 600$\,days
at 8\,GHz. 
Additional insight may be gained from future polarimetric studies
which have not been considered in this letter.
From all wavelet transforms of sources monitored as part of the UMRAO program, only for
OJ\,287 a comparably strong persistent signal has been found. The strength of the
modulation and the comparatively short time scale makes 
PKS\,B0048$-$097 a prime
object for coordinated broadband observing campaigns. In particular, the
correlation of the varying VLBI-jet ejection angle of the source with
radio (and higher energy) light-curve evolution provides a tool to test
precessing-jet models against alternative models 
(e.g., intrinsic instabilities of the flow)
of periodic light-curve modulation in blazars. 
Unlike most other sources,
the short 350\,day to 600\,day time scale of PKS\,B0048$-$097 allows these alternatives to be
tested observationally {in a relatively short time.}

\begin{acknowledgements}
We thank the referee, Claudia M. Raiteri, for her valuable suggestions on
improving the paper. 
MK has been supported in part by a Fellowship of the International Max
Planck Research School for Radio and Infrared Astronomy
at the Universities of Bonn and Cologne
and in part by an appointment to the NASA
Postdoctoral Program at the Goddard Space Flight Center, administered by
Oak Ridge Associated Universities through a contract with NASA.
PAH was supported in part by NSF grant AST 0205105.
UMRAO is partially supported  by a series of grants from the NSF and by funds
from the University of Michigan. 
The VLBA is operated by NRAO, which is a
facility of the National Science Foundation operated under
cooperative agreement by Associated Universities, Inc.
Part of this 
work has been made in collaboration with the VLBA 2\,cm Survey Team.
\end{acknowledgements}

\end{document}